\documentclass[12pt,draftclsnofoot, onecolumn]{IEEEtran}
\usepackage{graphicx,cite,epsfig,amssymb,amsmath,subfigure,url,stfloats,latexsym}
\usepackage{multirow,array}
\usepackage{graphicx,cite,amsmath,amssymb,hhline,booktabs,stfloats,bm}
\usepackage{verbatim}
\usepackage[utf8]{inputenc}
\usepackage{subfigure}
\usepackage{algorithm,algorithmic}
\usepackage{color}

\begin{document}
\setcounter{page}{1}
\markboth{IEEE Wireless Communications, Vol. XX, No. YY, Month 2017}
{Rong, Qian \& Chen: WiMAX \ldots}

\title{\mbox{}\vspace{1.5cm}\\
\textsc{
Optimal Task Allocation in Near-Far Computing Enhanced C-RAN for Wireless Big Data Processing} \vspace{1.5cm}
}

\author{Lianming Zhang, Kezhi Wang, \textit{IEEE Member}, Du Xuan and Kun Yang, \textit{IEEE Senior Member}.
\thanks{Lianming Zhang  (zlm@hunnu.edu.cn) is with College of Physics and Information Science, Hunan Normal University, Changsha, China,
Kezhi Wang (kezhi.wang@northumbria.ac.uk) is with
Department of Computer and Information Sciences, Northumbria University, NE2 1XE, Newcastle upon Tyne, U.K,
Xuan Du (xdua@essex.ac.uk) and Kun Yang (kunyang@essex.ac.uk) are with the School of Computer Sciences and Electrical Engineering, University of Essex, CO4 3SQ, Colchester, UK, Kun Yang is also with University of Electronic Science and Technology of China, Chengdu, China (Corresponding Author: Kun Yang).}
}

\date{\today}
\renewcommand{\baselinestretch}{1.2}
\thispagestyle{empty} \maketitle \thispagestyle{empty}

\begin{abstract}
With the increasing popularity of user equipments (UEs), the corresponding UEs' generating big data (UGBD) is also growing substantially, which makes both UEs and current network structures struggling in processing those data and applications. This paper proposes a Near-Far Computing Enhanced C-RAN (NFC-RAN) architecture, which can better process big data and its corresponding applications. NFC-RAN is composed of near edge computing (NEC) and far edge computing (FEC) units. NEC is located in remote radio head (RRH), which can fast respond to delay sensitive tasks from the UEs, while FEC sits next to baseband unit (BBU) pool which can do other computational intensive tasks. The task allocation between NEC or FEC is introduced in this paper. Also WiFi indoor positioning is illustrated as a case study of the proposed architecture. Moreover, simulation and experiment results are provided to show the effectiveness of the proposed task allocation and architecture.

\end{abstract}


\IEEEpeerreviewmaketitle

\vspace{0.3in}
\section{Introduction}

With the increasing popularity of user equipments (UEs) such as smartphones and hand-held devices, more and more resource-hungry applications like high definition video
gaming and virtual reality applications are developing and coming into play in our mobile devices. Due to limited resources in UEs, it is very difficult for them to compute the resource intensive applications. Moreover, UEs' generating big data (UGBD) is growing accordingly, which poses big challenges to the existing mobile devices and wireless networks \cite{7295483}.

Mobile edge computing (MEC) \cite{MEC} and offloading techniques \cite{6195845} have been proposed to enable UEs to send its tasks to their corresponding virtual machines. In this case, UEs’ experience will be increased substantially and their battery life will be prolonged largely. By taking advantage of the cloud technology, wireless networks have also undergone revolution recently \cite{7444125,7143336}. Cloud radio access network (C-RAN) has been proposed \cite{China} by moving most of the signal processing tasks, which previous were done in special hardware, now to the cloud, i.e., baseband unit (BBU) pool.
In this architecture, remote radio heads (RRHs) can be distributed easily according to the requirement. In C-RAN, we can dynamically and easily adjust and allocate computing resource to the wireless communications. Previous studies \cite{7511044, 7393804} have proposed to have mobile edge computing resource next to BBU pool to better provide service to the UEs. However, due to the transmission latency and limited bandwidth in fronthaul, the above architecture may be not beneficial to the delay sensitive tasks and big data applications.

Moreover, the studies in \cite{7562344} have investigated the big data applications in wireless communications and shown that it is not easy to process those data in wireless networks due to the required large amount of computation resource. Reference \cite{7909159} has proposed a big data computing architecture for
smart grid analytics. Some key technologies
to enable big-data-aware wireless communication
for smart grid were investigated in this paper. Reference \cite{7864795} has shown that
most of the applications require very short response time which is typically composed of two parts: transmission (i.e. communication) and processing (i.e., computation).

To better process big data applications, we propose a near-far computing enhanced C-RAN (NFC-RAN) architecture, by extending the C-RAN enhanced with mobile cloud \cite{7393804} with another layer of cloud computing, called near edge computing (NEC). In comparison with the mobile cloud in \cite{7393804}, which is referred to as far edge computing (FEC), NEC is deployed in the RRHs, namely much closer to UEs. Also, in this paper, we introduce how we allocate different tasks between NEC and FEC, as proper allocation affects the performance of the whole networks and the experiences of the user significantly \cite{6863135}.

Furthermore, in this paper, we will use the indoor positioning \cite {7442075,7841583} as an example to showcase the benefits of the proposed network architecture. To fulfil indoor positioning effectively, a large amount of wireless data concerning signal strength of positioning beacons needs to be collected, transmitted and processed. Also, unlike the outdoor positioning where the distance is usually measured in terms of tens of meters, indoor positioning techniques require to capture movement at a level of no more than 2-3 meters. This requires a very short response time when processing above mentioned large amount of information and the corresponding processing has to be conducted promptly in order to show a walker’s current position in a real-time manner.

The remainder of the paper is organized as follows. Section II describes the proposed NFC-RAN architecture on top of the popular C-RAN and presents various tasks involved in indoor positioning. Then the big data nature of indoor positioning problem is also discussed. Section III generalizes the task allocation issue in NFC-RAN into an optimization problem covering both the computation and communication aspects. Also, simulation result is given in this section. Section IV introduces the indoor positioning as a case study to illustrate our proposed architecture. Also, experimental results are reported in this section. Finally, conclusion remarks are given in Section V.

\vspace{0.3in}
\section{NFC-RAN Architecture and its Application to Indoor Positioning}

\subsection{NFC-RAN Architecture}

\begin{figure}[H]
\centering
\includegraphics[width=5.5in]{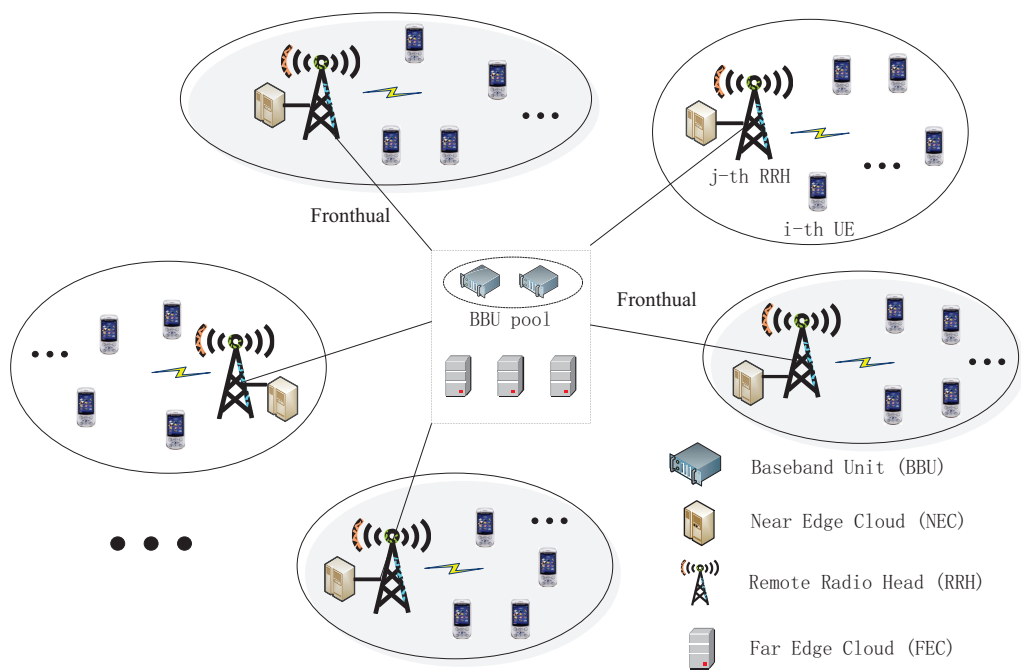}
\caption{Near-Far Edge Computing Enhanced C-RAN Architecture to Support Big Data Processing.} \label{c-ran}
\end{figure}

Our proposed NFC-RAN is shown in Fig. \ref{c-ran}. NFC-RAN is composed of the BBU pool, which is responsible of doing most of the signal processing tasks, and RRHs, which is in charge of sending and receiving data to and from UEs. RRHs can serve as the access points, which can be distributed closer to UEs as required. Also, RRHs is connected to the BBU pool through high speed fronthaul link. To support wireless big data processing, similar to \cite{7393804}, we propose to have FEC located next to BBU pool. If FEC decides to execute the task for UEs, UEs will send all the task data to RRHs through wireless channel first, and then RRHs will forward the data to FEC via fronthaul link. This may not be beneficial to the delay sensitive tasks or the tasks involving big transmission data. Thus, in addition to FEC, we propose to have the NEC located in each RRH as well. NEC can respond to UEs' requests much faster, due to its closer geographic location. In this architecture, UE does not have to send the data all the way to the central cloud, i.e., FEC.
This can not only save the bandwidth for fronthaul, but also reduce the response time for the tasks.

However, NEC may not have enough computational resource to process requests from all the UEs from its serving premises. Some delay tolerant tasks which require more computation can be forwarded to FEC instead. Thus, it is important to identify whether the tasks from UEs are delay tolerant or computation intensive or both and then allocate them to FEC and NEC accordingly. Next, we will analyze the big data tasks involved in indoor positioning.

\subsection{Tasks Involved in Indoor Positioning}
Outdoor positioning has been widely used in real life thanks to the Global Positioning System (GPS) technology \cite{7731599}. However, GPS does not work indoor whereas demands on finding indoor locations are high, as people spend most of their time indoor rather outdoor. Much effort has been made for indoor positioning, especially regarding techniques that do not involve much effort for initial deployment of positioning beacons. Due to the wide spread of WiFi hot spots (technically access points or APs), WiFi becomes a cost-effective option for indoor positioning. The WiFi assisted indoor positioning works in the steps as follows.

Firstly, WiFi AP’s signal fingerprints which are typically composed of received signal strength (RSS) are collected against a particular physical indoor location (also known as reference point). Then, after certain signal processing in cloud,
those (RSS, location) pairs
are stored in a database, which can be referred to as Signal Fingerprints Database (SFD).

Secondly, one can repeat the first step until all the designated locations are traversed. This procedure can be called as the site survey, which is conducted in an offline manner.

\begin{figure}[H]
\centering
\includegraphics[width=4.2in]{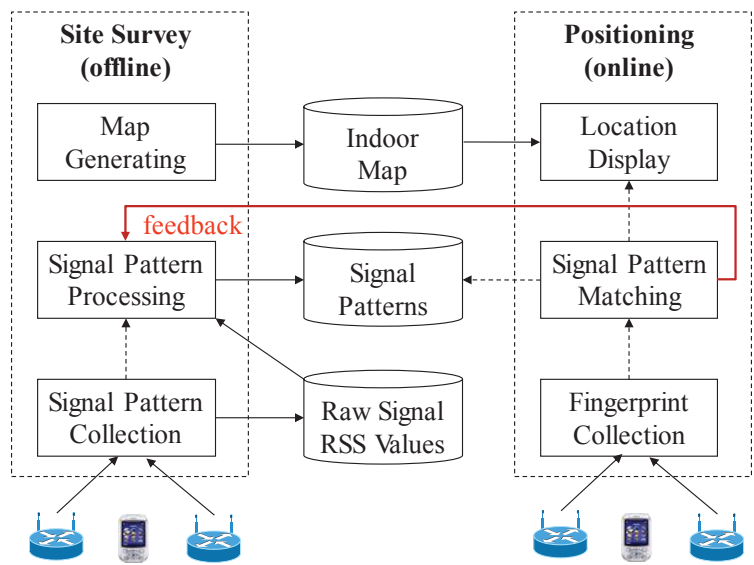}
\caption{The Software Architecture and the Corresponding Tasks in WiFi assisted Indoor Positioning.} \label{fig2}
\end{figure}

Then during the online positioning stage, the UE which requires to show its current location collects the current RSS of the WiFi signal at the surrounding area and then compare the RSS with the data in SFD. To this end, the location with the best-matched fingerprint can be considered as the UE’s current location. The software architecture and the corresponding tasks of this widely used fingerprint-based indoor positioning is depicted in Fig. \ref{fig2}.

To make positioning more accurate and efficient, extra procedures are introduced, such as signal pattern processing to clean out and correct the wrong fingerprints. Also, as RSS may change along with the change of indoor environment, to reflect in the signal pattern database on the positioning server, feedback is also applied in Fig. \ref{fig2}.

In summary there are two types of tasks. The first one is non-realtime or delay tolerant tasks that make long-term effect to the system such as signal pattern collection, processing and map generation, etc. The other type is realtime or delay sensitive tasks that are more concerned with end user’s positioning or display, such as instantaneous signal fingerprint collection, location display and signal pattern matching, etc. The non-realtime tasks can run remotely on FEC whereas the realtime tasks have to be executed on NEC to ensure fast response action. The experimental results in Section IV show the effectiveness of NEC in improving the accuracy of location.

\subsection{Wireless Big Data Involved in Indoor Positioning}

\begin{figure}[H]
\centering
\includegraphics[width=6.2in]{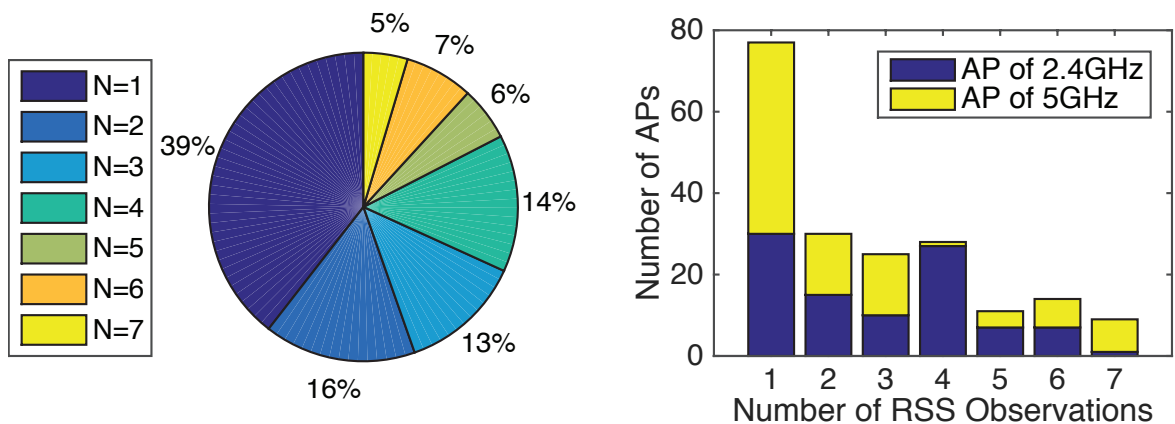}
\caption{Illustration of Wireless Big Data Involved in Indoor Positioning.} \label{fig3}
\end{figure}

To have an estimate of how many WiFi signals are needed for indoor positioning, certain tests are carried out in our experimental site, i.e., the 5th Floor of the Network Centre Building on the Colchester main campus of the University of Essex. The pie chart in Fig. \ref{fig3} shows the percentage of observed APs with different appearance frequency (denoted by $N$) when a UE is moving along a corridor for 25 seconds. Each AP is uniquely identified by a Basic Service Set Identifier (BSSID). Around 200 APs were detected during the 25 seconds’ movement. Around one third of APs are observed only once and about a quarter of APs are observed for more than 3 times. The AP appearing less times normally means its signal strength are weak and thus can be observed only within a certain range. The bar chart in Fig. \ref{fig3} illustrates the distribution of observed APs in 2.4 GHz and 5 GHz frequency band respectively. APs of 5 GHz dominates the observations with smallest and largest appearance frequency, which reveals that 5 GHz channels may be less crowded and weak signals in 5 GHz are more likely to be observed than the signals in 2.4 GHz.

Fig. \ref{fig3} also shows that approximately 600 observations are collected within 25 seconds, which means that about 25 APs are observed every second on average. For each AP, a large amount of information has to be collected, such as MAC address, RSS, working frequency and time stamp, etc, amounting to at least 8 bytes. Thus, a stream of data of more than 200 bytes per second is generated.
When the number of UEs or APs increases, the amount of data just for the purpose of indoor positioning will increase significantly. Note that this is a stream of data that is generated constantly and continually. Thus, it is nearly impossible for UEs to process those tasks with huge amount of data. The only way is to offload the corresponding tasks to the cloud, i.e., NEC and FEC.
Next, we will introduce how we allocate the tasks to NEC and FEC in our proposed architecture.

\section{Task Allocation in NFC-RAN for Wireless Big Data}

\subsection{System Model}
To better describe the task allocation algorithms, we assume that there are $M$ RRHs and each of which $j=1,2,...,M$ forms a small cell, which
can support $N_j$ UEs, as shown in Fig. \ref{c-ran}. Also, assume each UE employs the orthogonal channel to transmit its data and there is no interference between each other. We assume each UE is only served by its nearest RRH which is predefined by its geographical position and signal strength.

We denote UE $i=1,2,...,N_j$ in the coverage of $j$-th RRH as $ij$-th UE, which
has a task $U_{ij}=(F_{ij}, D_{ij},  T_{ij}), \forall i\in N_j, \forall j\in M$, where $F_{ij}$ (in cycles) describes the computation requirement of this task, $D_{ij}$ (in bits) denotes the data required to be transmitted to NEC, or FEC and $T_{ij}$ (in seconds) is the delay requirement in order to satisfy the UE's quality of service. We assume there is one FEC, next to BBU pool, with huge computation capacity $F^{FE}$ (in cycles/seconds) and it can be allocated to any UE in any cell. We also assume each RRH $j$ has a small NEC with limited computation capacity $F_j^{NE}$ (in cycles/seconds).

Also, we assume that the tasks cannot be executed in UEs, as UEs may not have enough processing capacity and thus UEs can either offload the tasks to FEC or NEC.
We define the indication parameters, i.e., $a_{ij}$, $b_{ij}$, $\forall i\in N_j, \forall j\in M$ to indicate where the task should be executed ($a_{ij}=1$ denotes the task is executed by FEC whereas $b_{ij}=1$ denotes the task is executed by $j$-th NEC). On the other hand, if $a_{ij}=b_{ij}=0$, it means this task can neither be executed by the NEC nor FEC and thus it has to be delayed to the next time slot.

If NEC decides to execute the task for $ij$-th UE, then it will allocate the CPU capacity $f_{ij}^{NE}$ to the UE, which needs to send
its data through wireless channel to $j$-th RRH with data rate $r_{ij}^W$.  In this case, the task with a large amount of data $D_{ij}$ (big data application) does not have to send all the data to the central cloud through the fronthaul link. On the other hand, if the task requires a huge amount of calculation (computation intensive application), NEC may not be able to complete this task, due to its limited computation capacity, i.e. $F_j^{NE}$. Then, UE has to send all the data to the central cloud, i.e., FEC.

If FEC decides to execute the task for UE, then it will allocate the CPU capacity $f_{ij}^{FE}$ to the $ij $-th UE. In this case, UE should first send its data to $j$-th RRH with wireless data rate $r_{ij}^W$, and then RRH will forward the data to FEC with fronthual transmission data rate as $r_{ij}^F$. Also, as the computation capacity of FEC is not infinite, i.e. constrained by the capacity of the physical machine, i.e., $F^{FE}$, some UEs may still not be able to complete the tasks. Moreover, if UEs decide to send the task to FEC, the capacity of fronthaul has to be taken into account. Thus, we assume the capability of the $j$-th frontaul as $R_j $.

We model the task allocation problem as follows.
\begin{equation}\label{en25}
\begin{aligned}
\mathcal{P}: \;\;\;& \underset{a_{ij}, b_{ij}
}{\text{max}}\;\;\;  \sum_{i\in N_j}\sum_{j\in M}( a_{ij} + b_{ij})
\\& \text{subject to }: C1: a_{ij} \left(\frac{D_{ij}}{r_{ij}^W} +\frac{D_{ij}}{r_{ij}^F} +\frac{F_{ij}}{f_{ij}^{FE}}\right )\leq T_{ij}
\\& C2: b_{ij} \left(\frac{D_{ij}}{r_{ij}^W} +\frac{F_{ij}}{f_{ij}^{NE}}\right )\leq T_{ij}
\\& C3: \sum_{i\in N_j}\sum_{j\in M} a_{ij} f_{ij}^{FE} \leq  F^{FE}
\\& C4: \sum_{i\in N_j} b_{ij} f_{ij}^{NE} \leq   F_j^{NE}
\\& C5: \sum_{i\in N_j} a_{ij} r_{ij}^{F} \leq   R_j
\\& C6: a_{ij} + b_{ij} \leq 1
\\& C7: a_{ij}, b_{ij} =\{ 0,1\}, \forall i\in N_j, \forall j\in M
\end{aligned}
\end{equation}
where we aim to maximize the successful rate of all the offloading tasks, by deciding where the tasks should be executed. In another words, we try to accommodate as many tasks in cloud as possible. In above problem, $C1$ and $C2$ denote that the task has to be completed in certain amount of time by FEC or NEC, respectively, $C3$ and $C4$ denotes that the computation resources are limited in FFC and $j$-th NEC respectively, $C5$ is the constraint for the $j$-th fronthaul, $C6$ and $C7$ can not only show where each task should be executed, but also make the problem feasible. The above problem may be modified as the multi-dimension multi-choice 0-1 knapsack
problem (MMKP), which can be solved effectively by using heuristic algorithm.

\subsection{Simulation Result}

\begin{figure}[H]
\centering
\includegraphics[width=6.6in]{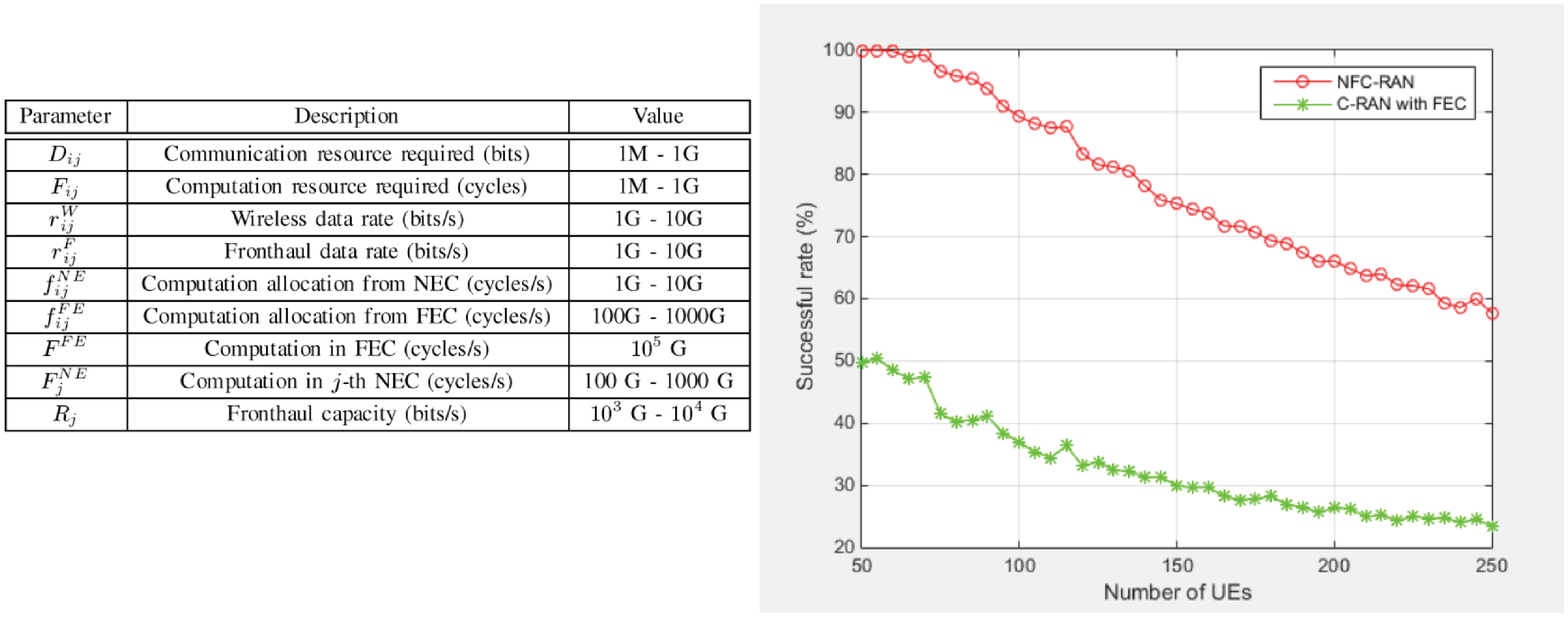}
\caption{Simulation Parameters Setting (Left) and Simulation Result (Right).} \label{figtable}
\end{figure}

We assume that there are five $M=5$ RRHs, each of which forms a small cell. In each cell, there are $N$ UEs, each of which has a task to be completed. For each task, we assume the latency requirement is 3 seconds. We assume that other parameters are randomly assigned from the sets indicated in left hand side of Fig. \ref{figtable}.

In the right hand side of Fig. \ref{figtable}, we show that the relation between task successful rate versus the number of offloading tasks.
The task successful rate or the task completion rate is defined as the ratio of the number of completion to the overall offloaded tasks.
We compare the new NFC-RAN and the traditional C-RAN architecture only with FEC. The number of UEs is set from 10 to 50 in each of the 5 cells. Also, to compare fairly, $F_{FE}$ is set to $10^7$ GHz for traditional C-RAN.

One can see that with the increase of the number of the offloading tasks, the successful rate decreases. This is because the cloud has limited computation resource and some tasks may be dropped or delayed to the next time interval. Our proposed NFC-RAN outperforms the traditional C-RAN with FEC, as NFC-RAN not only supports FEC, but also NEC which is much closer to UEs. This structure is beneficial to the delay sensitive tasks and therefore can increase the overall tasks' successful rate. In the next section, we will use indoor positioning as a case study to show the benefit of NFC-RAN.

\section{Case Study: Indoor Positioning}
This section use indoor positioning as a case study and we assume that the task allocation is predetermined, namely, all the offline tasks and the feedback tasks (refer to Fig. \ref{fig2}) are executed at FEC whereas the online computation tasks are executed at NEC. The experiments are carried out to show the performance improvement of positioning by using NFC-RAN architecture.

\subsection{Experimental Setup}

\begin{figure}[H]
\centering
\includegraphics[width=5in]{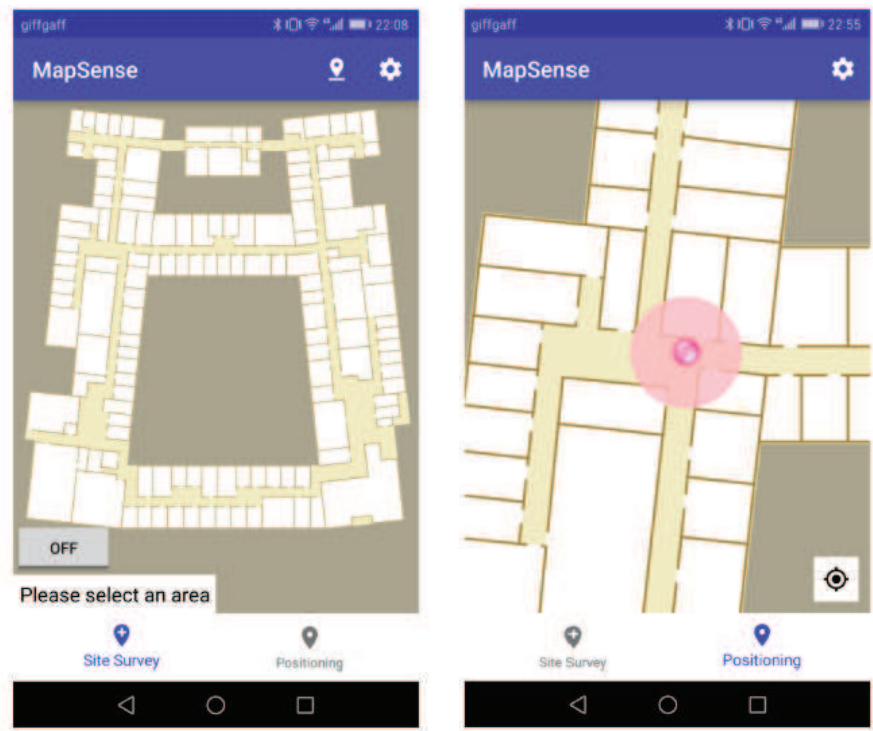}
\caption{Illustration of Indoor Map of the Experimental Site (Left) and Positioning Result (Right) Shown on Android Smartphone.} \label{fig4}
\end{figure}
We setup our testbed on the 5th Floor of the Network Centre Building of the University of Essex, where the experiments were conducted. The map of the environment is depicted in left hand side of Fig. \ref{fig4}. The site is covered by about 60 wireless APs of Aruba mounted on the ceiling. They are part of the campus WiFi infrastructure. Each physical AP may generate multiple SSIDs, which means much more APs are observed by UEs. The circle in right hand side of Fig. \ref{fig4} indicates the location of the UE.

\subsection{Experimental Results}

\begin{figure}[H]
\centering \subfigure[]{\includegraphics [width=4.5in]{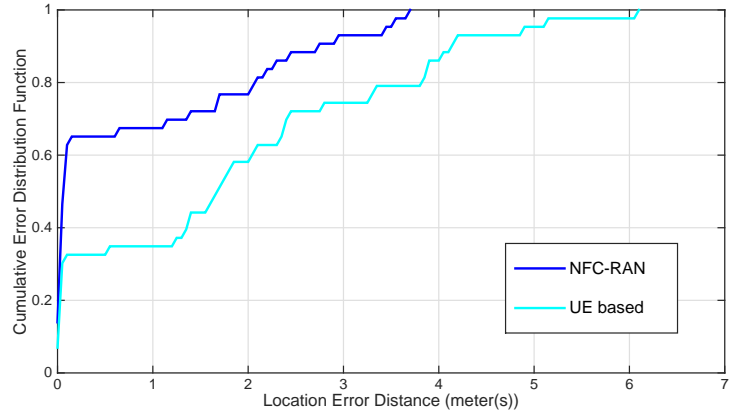}}
\subfigure[]{\includegraphics [width=4.5in]{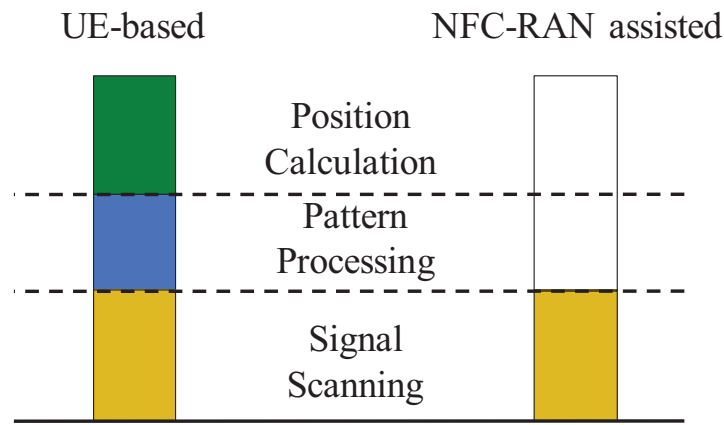}}
\caption{Positioning Accuracy and Energy Saving on UE, (a) Comparison of CDF Between NFC-RAN Based and UE Based System, (b) Overall Energy Consumption on UE. } \label{wired}
\end{figure}

In this section, two experiments were conducted. In the first one, most of the online tasks such as fingerprint matching are offloaded to and conducted by NFC-RAN, whereas in the second experiment, UE itself executes most of the task. We compare these two cases in Fig. \ref{wired} where in Fig. \ref{wired} (a), we show the cumulative distribution function (CDF) of location errors for both cases.

One can see that our proposed NFC-RAN assisted system shows better location accuracy than UE-based one. The percentage of accuracy within 1 meter is approximately 70\% in the NFC-RAN assisted system, which doubles the percentage of UE-based approach (35\%). In 90\% of positioning results, the NFC-RAN assisted and UE-based systems provide accuracy of 3 meters and 5 meters respectively. Through the comparison of the location errors when the CDF is 100\%, it is apparent that the maximum error distance of UE-based system is almost 7.5 meters whereas the NFC-RAN assisted system can decrease it to 4.5 meters. The major reasons why NFC-RAN architecture outperforms UE based one are two-fold. Firstly, the processing capacity of UE is in no comparison with that of NFC-RAN system which is composed of much more powerful computation resource. The high computation capacity in NFC-RAN can process more patterns (big data applications) and to execute more comprehensive tasks. Secondly, the proposal of NEC, which brings computation closer to the UE, can fast calculate and execute tasks for the UE, such as signal pattern matching tasks. However, UE based positioning approach may lead to a situation where the produced location result by UE is out of date sometimes, thus resulting in a worse location accuracy.

In Fig. \ref{wired} (b), we show that the overall energy consumption on UE in the situation where NFC-RAN dealing with computation or UE itself conducting computation. From Fig. \ref{wired} (b), we can see that UE will save a lot of energy on the NFC-RAN assisted situation, as most of the tasks are offloaded from UE to NFC-RAN. This is particularly important for the practical indoor positioning system as the positioning process may incur a large amount of data and therefore drain the UE’s battery quickly.

\section{Conclusions}
In this paper, we have proposed NFC-RAN architecture, which can facilitate wireless big data processing. NFC-RAN is composed of NEC and FEC, where FEC is located next to BBU, which can provide large amount of computational resource to UEs, while NEC, located in RRH, can fast respond to the delay sensitive applications. Also, task allocation in NFC-RAN for wireless big data is illustrated and indoor positioning, as a case study, is exemplified to show the benefit of the proposed architecture. Future work will focus on how to allocate and execute the tasks dynamically, including in the NEC, FEC and UE itself. Also, more efficient task allocation algorithms for wireless big data will be investigated.

\vspace{0.3in}
\section{Acknowledgements}
This work was supported in part by Natural Science Foundation of China (Grant No. 61620106011, 61572389 and 61572191), UK EPSRC NIRVANA project (EP/L026031/1), EU Horizon 2020 iCIRRUS project (GA-644526) and EU FP7 Project CROWN (GA-2013-610524). We also would like to thank Dr Yuansheng Luo for the very useful discussion.

\vspace{0.3in}
\bibliographystyle{ieeetran}
\bibliography{bare_jrnl}

\begin{thebibliography}{10}
\providecommand{\url}[1]{#1}
\csname url@samestyle\endcsname
\providecommand{\newblock}{\relax}
\providecommand{\bibinfo}[2]{#2}
\providecommand{\BIBentrySTDinterwordspacing}{\spaceskip=0pt\relax}
\providecommand{\BIBentryALTinterwordstretchfactor}{4}
\providecommand{\BIBentryALTinterwordspacing}{\spaceskip=\fontdimen2\font plus
\BIBentryALTinterwordstretchfactor\fontdimen3\font minus
  \fontdimen4\font\relax}
\providecommand{\BIBforeignlanguage}[2]{{%
\expandafter\ifx\csname l@#1\endcsname\relax
\typeout{** WARNING: IEEEtran.bst: No hyphenation pattern has been}%
\typeout{** loaded for the language `#1'. Using the pattern for}%
\typeout{** the default language instead.}%
\else
\language=\csname l@#1\endcsname
\fi
#2}}
\providecommand{\BIBdecl}{\relax}
\BIBdecl

\bibitem{7295483}
S.~Bi, R.~Zhang, Z.~Ding, and S.~Cui, ``Wireless communications in the era of
  big data,'' \emph{IEEE Communications Magazine}, vol.~53, no.~10, pp.
  190--199, October 2015.

\bibitem{MEC}
M.~Patel, B.~Naughton, C.~Chan, N.~Sprecher, S.~Abeta, and A.~Neal, ``Mobile
  edge computing white paper,'' \emph{ETSI}, 2014.

\bibitem{6195845}
S.~Kosta, A.~Aucinas, P.~Hui, R.~Mortier, and X.~Zhang, ``Thinkair: Dynamic
  resource allocation and parallel execution in the cloud for mobile code
  offloading,'' in \emph{2012 IEEE Proceedings INFOCOM}, March 2012, pp.
  945--953.

\bibitem{7444125}
M.~Peng, Y.~Sun, X.~Li, Z.~Mao, and C.~Wang, ``Recent advances in cloud radio
  access networks: System architectures, key techniques, and open issues,''
  \emph{IEEE Communications Surveys Tutorials}, vol.~18, no.~3, pp. 2282--2308,
  thirdquarter 2016.

\bibitem{7143336}
J.~Wu, D.~Liu, X.~L. Huang, C.~Luo, H.~Cui, and F.~Wu, ``{DaC-RAN}: A
  data-assisted cloud radio access network for visual communications,''
  \emph{IEEE Wireless Communications}, vol.~22, no.~3, pp. 130--136, June 2015.

\bibitem{China}
{China Mobile Research Institute}., ``{C-RAN} white paper: The road towards
  green {Ran}. [online],'' \emph{(Jun. 2014)}, Available:
  http://labs.chinamobile. com/cran.

\bibitem{7511044}
K.~Wang, K.~Yang, X.~Wang, and C.~S. Magurawalage, ``Cost-effective resource
  allocation in {C-RAN} with mobile cloud,'' in \emph{2016 IEEE International
  Conference on Communications (ICC)}, May 2016, pp. 1--6.

\bibitem{7393804}
K.~Wang, K.~Yang, and C.~Magurawalage, ``Joint energy minimization and resource
  allocation in {C-RAN} with mobile cloud,'' \emph{IEEE Transactions on Cloud
  Computing}, vol.~PP, no.~99, pp. 1--1, 2016.

\bibitem{7562344}
X.~Zhang, Z.~Yi, Z.~Yan, G.~Min, W.~Wang, A.~Elmokashfi, S.~Maharjan, and
  Y.~Zhang, ``Social computing for mobile big data,'' \emph{Computer}, vol.~49,
  no.~9, pp. 86--90, Sept 2016.

\bibitem{7909159}
K.~Wang, Y.~Wang, X.~Hu, Y.~Sun, D.~J. Deng, A.~Vinel, and Y.~Zhang, ``Wireless
  big data computing in smart grid,'' \emph{IEEE Wireless Communications},
  vol.~24, no.~2, pp. 58--64, April 2017.

\bibitem{7864795}
K.~Wang, K.~Yang, H.~H. Chen, and L.~Zhang, ``Computation diversity in emerging
  networking paradigms,'' \emph{IEEE Wireless Communications}, vol.~24, no.~1,
  pp. 88--94, February 2017.

\bibitem{6863135}
G.~Nan, Z.~Mao, M.~Li, Y.~Zhang, S.~Gjessing, H.~Wang, and M.~Guizani,
  ``Distributed resource allocation in cloud-based wireless multimedia social
  networks,'' \emph{IEEE Network}, vol.~28, no.~4, pp. 74--80, July 2014.

\bibitem{7442075}
X.~Du and K.~Yang, ``A map-assisted wifi ap placement algorithm enabling mobile
  device indoor positioning,'' \emph{IEEE Systems Journal}, vol.~PP, no.~99,
  pp. 1--9, 2016.

\bibitem{7841583}
X.~Du, J.~Wu, K.~Yang, and L.~Wang, ``An ap-centred indoor positioning system
  combining fingerprint technique,'' in \emph{2016 IEEE Global Communications
  Conference (GLOBECOM)}, Dec 2016, pp. 1--6.

\bibitem{7731599}
G.~Xu, S.~Gao, M.~Daneshmand, C.~Wang, and Y.~Liu, ``A survey for mobility big
  data analytics for geolocation prediction,'' \emph{IEEE Wireless
  Communications}, vol.~24, no.~1, pp. 111--119, February 2017.

\end{thebibliography}

\end{document}